\documentclass[journal]{IEEEtran}

\usepackage[T1]{fontenc}
\usepackage{graphicx}
\usepackage[caption=false]{subfig}
\usepackage{mathrsfs,amsmath,amsfonts,amssymb,amsthm,amstext,amscd,amsxtra,amsopn}
\usepackage{diagbox}
\usepackage{lettrine}
\usepackage{cite}
\usepackage{eucal}
\providecommand{\keywords}[1]{\textbf{\textit{Index terms---}} #1}
\usepackage{comment}
\usepackage{bm}
\usepackage{multirow}
\usepackage{esint}
\usepackage{empheq}
\usepackage{bm}
\usepackage{algorithm}
\usepackage[noend]{algpseudocode}
\usepackage{etoolbox}
\usepackage{leftidx}
\usepackage{makecell}
\usepackage{stackrel}
\usepackage{xcolor}
\usepackage{comment}
\usepackage{nicefrac}
\usepackage[normalem]{ulem}
\usepackage[hyphens]{url}
\usepackage[hidelinks]{hyperref}
\hypersetup{breaklinks=true}
\renewcommand{\vec}[1]{\bm{#1}}

\hypersetup{breaklinks=true}
\DeclarePairedDelimiter\abs{\lvert}{\rvert}
\newcommand\norm[1]{\left\lVert#1\right\rVert}

\newcommand\It{{\cal{I}}}
\newcommand\Kt{{\cal{K}}}

\newcommand\Mt{{\cal{M}}}
\newcommand\Nt{{\cal{N}}}

\makeatletter
\def\BState{\State\hskip-\ALG@thistlm}

\expandafter\patchcmd\csname\string\algorithmic\endcsname{\itemsep\z@}{\itemsep=1.5mm plus2pt}{}{}

\makeatother

\algnewcommand{\Initialize}[1]{%
  \State \textbf{Initialize:}
  \Statex \hspace*{\algorithmicindent}\parbox[t]{.8\linewidth}{\raggedright #1}
}

\newcommand{\iu}{{\mathrm{i}}}

\title{Magnetic-resonance-based electrical property mapping using Global Maxwell Tomography with an 8-channel head coil at 7 Tesla: a simulation study}

\author{Ilias I. Giannakopoulos, ~\IEEEmembership{Member,~IEEE},\\ Jos{\'e} E.C. Serrall{\'e}s, ~\IEEEmembership{Member,~IEEE}, Luca Daniel, ~\IEEEmembership{Member,~IEEE}, Daniel K. Sodickson,\\ Athanasios G. Polimeridis, ~\IEEEmembership{Senior Member,~IEEE}, Jacob K. White, ~\IEEEmembership{Fellow,~IEEE},\\ Riccardo Lattanzi, ~\IEEEmembership{Senior Member,~IEEE}

\thanks{This work was supported in part by the Skoltech-MIT Next Generation Program, the National Science Foundation (NSF 1453675) and the National Institutes of Health (NIH R01EB024536, NIH P41EB017183). (Corresponding author: Ilias I. Giannakopoulos.)}
\thanks{Ilias I. Giannakopoulos is with the Skoltech Center for Computational Data-Intensive Science and Engineering, Skolkovo Institute of Science and Technology, 143026 Moscow, Russia, and the Research Laboratory of Electronics, Department of Electrical Engineering and Computer Science, Massachusetts Institute of Technology, Cambridge, MA, USA.}
\thanks{Jos{\'e} E.C. Serrall{\'e}s, Luca Daniel and Jacob K. White are with the Research Laboratory of Electronics, Department of Electrical Engineering and Computer Science, Massachusetts Institute of Technology, Cambridge, MA, USA.}
\thanks{Daniel K. Sodickson and Riccardo Lattanzi are with the Center for Advanced Imaging Innovation and Research (CAI$^{\text{2}}$R) and the Bernard and Irene Schwartz Center for Biomedical Imaging (CBI), Department of Radiology, New York University School of Medicine, New York, NY, USA)}
\thanks{Athanasios G. Polimeridis is with Q Bio, CA 94063, USA.}}

\begin{document}
\bstctlcite{IEEEexample:BSTcontrol}

\maketitle

\begin{abstract} 
\textit{Objective:} Global Maxwell Tomography (GMT) is a recently introduced volumetric technique for noninvasive estimation of electrical properties (EP) from magnetic resonance measurements. Previous work evaluated GMT using ideal radiofrequency (RF) excitations. The aim of this simulation study was to assess GMT performance with a realistic RF coil. \textit{Methods:} We designed a transmit-receive RF coil with $8$ decoupled channels for $7$T head imaging. We calculated the RF transmit field ($B_1^+$) inside heterogeneous head models for different RF shimming approaches, and used them as input for GMT to reconstruct EP for all voxels. \textit{Results:} Coil tuning/decoupling remained relatively stable when the coil was loaded with different head models. Mean error in EP estimation changed from $7.5\%$ to $9.5\%$ and from $4.8\%$ to $7.2\%$ for relative permittivity and conductivity, respectively, when changing head model without re-tuning the coil. Results slightly improved when an SVD-based RF shimming algorithm was applied, in place of excitation with one coil at a time. Despite errors in EP, RF transmit field ($B_1^+$) and absorbed power could be predicted with less than $0.5\%$ error over the entire head. GMT could accurately detect a numerically inserted tumor. \textit{Conclusion:} This work demonstrates that GMT can reliably reconstruct EP in realistic simulated scenarios using a tailored 8-channel RF coil design at $7$T. Future work will focus on construction of the coil and optimization of GMT's robustness to noise, to enable in-vivo GMT experiments. \textit{Significance:} GMT could provide accurate estimations of tissue EP, which could be used as biomarkers and could enable patient-specific estimation of RF power deposition, which is an unsolved problem for ultra-high-field magnetic resonance imaging.
\end{abstract}


\section{Introduction} \label{sc:I}

\IEEEPARstart{M}{agnetic} resonance (MR) imaging (MRI) has become an indispensable and powerful tool for disease diagnosis and characterization of pathologies. $7$ Tesla ($7$ T) MRI scanners, which have recently been introduced clinically \cite{siemens2017seventesla}, can provide detailed images of the interior structure of the human body, by exploiting the larger signal-to-noise ratio (SNR) available with the higher static magnetic field as compared with lower field strengths such as $1.5$ T or $3$ T, which have historically been used for clinical imaging. However, the higher operating frequency of these scanners corresponds to a shorter wavelength that yields strong interactions between electromagnetic (EM) fields and biological tissue. The resulting interference patterns can degrade image quality and cause local amplifications of radiofrequency (RF) power deposition, which is a safety concern. For these reasons, accurate modeling of RF transmit coils and the associated EM fields has become critical for practical applications of ultra-high-field ($\geq 7$ T) MRI, which seek to maximize performance without compromising patient safety \cite{Lattanzi2009, jin1997sar, zhang2014quantitative, zhang2013complex, cosottini2014short}. However, one can only perform a limited number of EM simulations with the available numerical human body models \cite{VirtualFamily}, which could lead to erroneous predictions of the EM field behavior in actual patients \cite{alon2016effects}.
\par
The capability of mapping tissue electrical properties (EP) noninvasively could enable the generation of more realistic human models, and could facilitate patient-specific simulations to calibrate optimized MRI examinations. Although ex-vivo or in-situ animal measurements have been made over time \cite{gabriel1996dielectricii}, experimental access to in-vivo tissue EP distributions has remained extremely limited. A variety of noninvasive MR-based EP estimation techniques have been proposed over the years, and these can be divided into two classes. One category uses MR measurements to directly compute EP, by solving a system of equations derived from the differential form of Maxwell's equations. These differential approaches \cite{katscher2009determination, gurler2017gradient, song2013grad2d, lee2015zet, marques2015b1m, liu2013txrx, zhang2013multiTx, hafalir2014convection, sodickson2012local} require first- and second-order numerical derivatives of noisy MR data, which result in unavoidable artifacts at boundaries between tissue regions with distinct EP values. Both of these effects limit the accuracy and achievable spatial resolution  of differential methods \cite{mandija2018error}. A second category relies on the integral form of Maxwell's equations and aims to extract EP by solving a large ill-conditioned inverse problem, in which MR measurements provide the data consistency term \cite{serralles2019noninvasive, balidemaj2015csiept, guo2017iept, hong2017vie, schmidt2016bie}. These integral approaches can be applied at any spatial resolution, although their computational complexity can become intractable for a large number of voxels, and their accuracy depends heavily on the regularization strategy. 
\par
Global Maxwell Tomography (GMT) is a recently proposed technique that belongs to the integral equation type \cite{serralles2019noninvasive}. GMT differs from the other integral methods because it is inherently volumetric and does not require any simplifying geometrical assumptions. GMT uses an iterative non-linear least squares optimization to estimate the EP from multiple measurements of the absolute value and relative phase of the transmit field $B_1^+$. It is accurate because it relies on higher-order volume integral equations (VIE) \cite{georgakis2019fast} for rapid and reliable solutions of the forward problem. GMT was first assessed in simulation with realistic human body models (RHBM), using the first eight eigenvectors of a numerical EM basis as hypothetical MR transmitters, to investigate the technique independently from particular coil designs \cite{serralles2019noninvasive}. In fact, since the numerical conditioning of GMT strongly depends on the number, distinctness and spatial distribution of the measured $B_1^+$ fields, the use of such a basis represents an ideal situation in which the incident fields generated for the excitation of the RHBM are highly inhomogeneous, distributed over the entire volume of interest and mutually orthogonal. More recently, GMT was evaluated in a numerical experiment using simulated incident fields from an $8$-element transmit-receive array, which led to accurate reconstructions of the EP of a four-compartment tissue-mimicking phantom \cite{giannakopoulos2019global}. GMT was also successfully demonstrated in an actual experiment at $7$ T using an $8$-element coil \cite{chen2018highly} and a homogeneous cylindrical phantom \cite{serralles2019noninvasive}.
\par
The aforementioned results showed that GMT can be accurate and robust not only with an ideal basis of incident EM fields, but also when realistic RF coils are used. However, GMT performance depends heavily on the characteristics of the EM fields generated by the RF coil inside the sample and, therefore, this preliminary work with simple phantom geometries is not sufficient to predict the feasibility of in-vivo experiments. The aim of the current work was to extend the work presented in \cite{giannakopoulos2019global} to design a novel problem-dedicated $8$-channel $7$ T head coil and investigate its performance in simulated GMT experiments with realistic anatomical models.
\par
This paper begins by presenting the technical background of GMT and the integral equation formulation used for EM modeling. It then describes the design of the RF coil array, analyzing its excitation patterns for different RF shimming configurations. Next, the sensitivity of GMT with respect to tuning, matching and decoupling of the array elements when loaded with different heterogeneous head models is evaluated. Finally, numerical GMT experiments are presented and discussed.

\section{Global Maxwell Tomography}  \label{sc:II}

GMT solves an inverse scattering problem to retrieve EP, given measured transmit magnetic field patterns ($\hat{B}_1^+$) inside the sample, a set of incident EM fields and an initial guess of the sample EP. The solution is obtained through a gradient-based optimization procedure, where in each iteration a set of $B_1^+$ fields is calculated from the incident EM fields and an updated guess of EP, and those calculated fields are compared with the measured $\hat{B}_1^+$. The estimation of the $B_1^+$ (the forward problem) is performed using the method of moments technique, where the robust current-based VIE (JVIE) \cite{Polimeridis2014} is discretized over a uniform grid, in order to significantly accelerate the matrix-vector products with the help of the fast Fourier transform (FFT) \cite{Borup1984}. In this work we employed piecewise linear basis functions, which have recently been shown to provide accurate EM fields calculations in MRI applications \cite{georgakis2019fast}. The solution of the forward problem starts by computing the polarisation currents in the scatter ($\vec{j}^b$) from the incident electric fields ($\vec{e}_{\text{inc}}$) and an estimated EP distribution ($\epsilon$) :
\begin{equation}
\left(\It -\Mt_{\frac{\epsilon - 1}{\epsilon}} \Nt \right) \vec{j}^b \left(\vec{r}\right) = \mathrm{i} \omega \epsilon_0 \Mt_{\frac{\epsilon - 1}{\epsilon}} \vec{e}_{\text{inc}}.
\end{equation}
Here $\iu$ is the imaginary unit, $\It$ is the identity operator, $\omega$ is the angular frequency, $\epsilon_0$ is the free space permittivity, $\Mt_{\frac{\epsilon - 1}{\epsilon}}$ is a multiplicative operator over the quantity in the subscript, $\Nt$ is the operator defined below and $\epsilon$ is the complex relative permittivity given by
\begin{equation}
\epsilon = \epsilon_r + \frac{\sigma_e}{\mathrm{i} \omega \epsilon_0},
\end{equation}
where $\epsilon_r$ is the scalar relative permittivity and $\sigma_e$ is the electric conductivity. The operators $\Nt$ and $\Kt$ map the polarisation currents $\vec{j}^b$ to the electric field $\vec{e}$ and  magnetic field $\vec{h}$, respectively, inside the scatterer:
\begin{equation}
\begin{aligned}
\Nt \left(\vec{j}^b\right) & \triangleq \nabla \times \nabla \times \int\limits_{\Omega} g\left(\vec{r}-\vec{r}'\right) \vec{j}^b\left(\vec{r}'\right)d^3\vec{r}' \\
\Kt \left(\vec{j}^b\right) & \triangleq \nabla \times \int\limits_{\Omega} g\left(\vec{r}-\vec{r}'\right) \vec{j}^b\left(\vec{r}'\right)d^3\vec{r}'.
\end{aligned}
\end{equation} 
$g\left(\vec{r}-\vec{r}'\right) \triangleq e^{-\iu k_0 \abs{\vec{r}-\vec{r}'}} / 4 \pi \abs{\vec{r}-\vec{r}'}$ is the free-space Green's function that models the interaction between the observation point and the source point, $\vec{r}$ and $\vec{r}'$, and $k_0$ is the wavenumber in free-space. Finally, $B_1^+$ is defined as
\begin{equation}
B_1^+ = \mu_0 \left(\vec{h}_x+\iu \vec{h}_y \right),
\end{equation}
with $\mu_0$ being the magnetic permeability of free space and the subscript of $\vec{h}$ denoting the coresponding Cartesian component.
\par 
The cost function of GMT is the weighted difference between measured ($\hat{B}_{1}^+$) and estimated ($B_{1}^+$) transmit fields \cite{serralles2019noninvasive}. Specifically, it is given by the following equation (using Einstein notation for summation over repeated indices)
\begin{equation}
\begin{aligned}
f(\epsilon_r,\sigma_e) &= \frac{\sqrt{\norm{ w_i \circ w_j \circ \left( \hat{B}_{1,i}^+ \circ \overline{\hat{B}_{1,j}^+} - B_{1,i}^+ \circ \overline{B_{1,j}^+} \right)}_2^2}}{\sqrt{\norm{ w_i \circ w_j  \circ \left( \hat{B}_{1,i}^+ \circ \overline{\hat{B}_{1,j}^+} \right) }_2^2}} + \\ &+ f_r(\alpha,\beta,\gamma).
\end{aligned}
\label{eq:nf}
\end{equation}
The symbol $\circ$ denotes the Hadamard product. The two summations over $j$ and $i$ iterate over all of the unique field maps of a multiple-channel transmit array, $w_i$ and $w_j$ are a set of weights chosen precisely to prioritize regions with higher SNR during the optimization procedure. Note that the cost function operates on the product of complex-valued fields and complex-conjugated fields (indicated by horizontal bars) in pairs of channels. This corresponds to the product of absolute values of $B_1^+$ maps for any two channels multiplied by a phase factor whose exponent is the relative phase between the two fields at all spatial positions. This relative phase can be extracted without any assumptions from standard MRI experiments, whereas the absolute phase of the transmit fields cannot \cite{sodickson2012local}. The term $f_r$ is the Match Regularization term \cite{serralles2019noninvasive} and it is included in the cost function to address additive physical noise in the measurements. The performance of the regularizer depends on three parameters $\{\alpha,\beta,\gamma\}$ that can be set for each problem independently. For additional information on the regularizer we refer the reader to \cite{serralles2019noninvasive}.

\section{Triangular Head RF-coil}  \label{sc:III}

GMT relies on solving an inverse problem using the complementary information provided by the $B_1^+$ maps of each coil \cite{serralles2017investigation}. Therefore, the numerical conditioning of GMT, and, consequently, its accuracy, depends on how distinct, or orthogonal, are the coil's $B_1^+$ fields among each other. For this reason, the feasibility of GMT was first assessed using eight perfectly orthogonal incident field vectors generated from a numerical EM basis \cite{serralles2019noninvasive}. However, these ideal excitations are challenging to generate with actual RF coils. To investigate GMT performance in a more realistic situation, we designed an $8$-element transmit-receive array (Fig. 1) and loaded it with a heterogeneous human head model (``Duke'', from the Virtual family population \cite{VirtualFamily}).

\renewcommand{\thefigure}{1}
\begin{figure}[ht!]
\begin{center}
\includegraphics[width=0.48\textwidth, trim={0 0 0 0}]{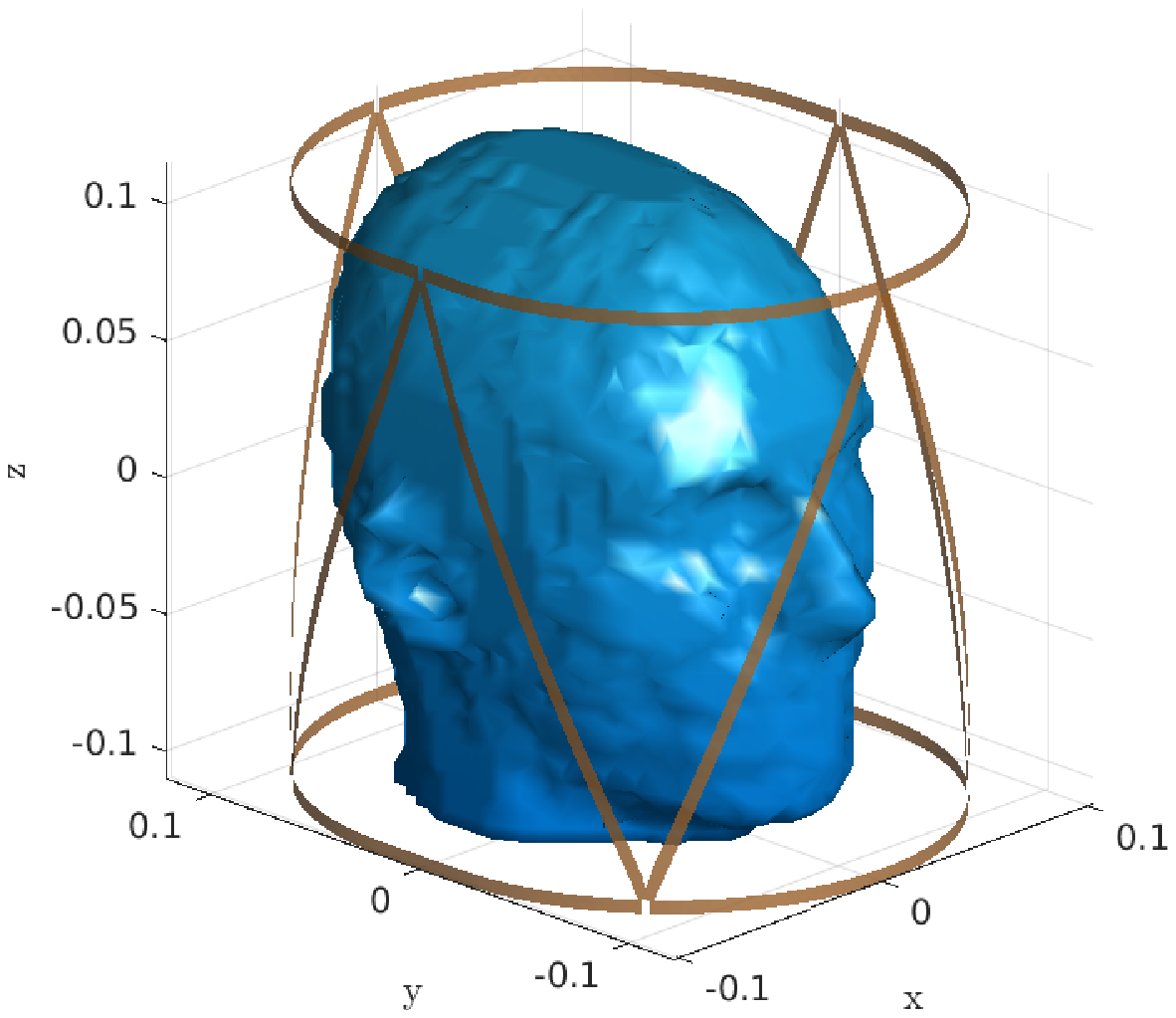}
\includegraphics[width=0.40\textwidth, trim={0 0 0 0}]{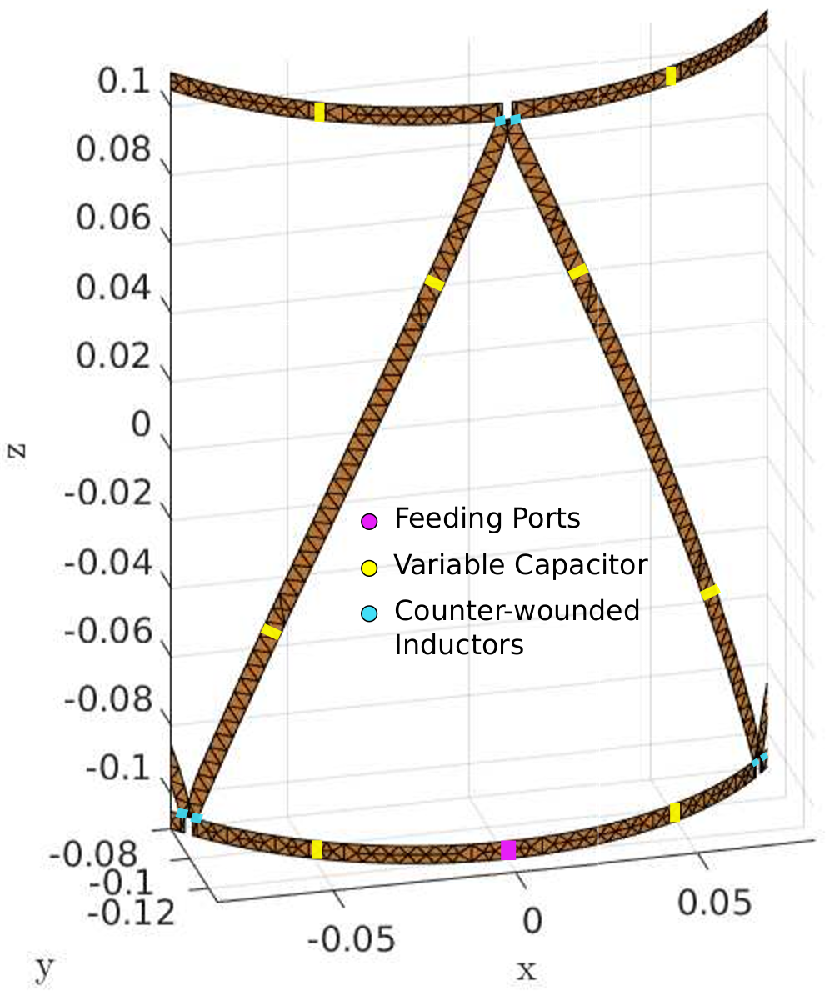}
\caption{Coil geometry. (top) The RF coil loaded with the realistic human head model ``Duke'', and (bottom) the feeding port and the lumped elements for one representative coil element.}
\label{fig:n1}
\end{center}
\end{figure}

\par
The coil's geometry, closely fitting the head model (with a gap ranging from $5$ mm to $20$ mm), is a stadium (discorectangle) with the radius of semicircles equal to $10.2$ cm, the length of the edge of the rectangle equal to $3.8$ cm and the height equal to $22$ cm. In the bottom of Fig. 1, the location of the lumped elements and the feeding ports are shown for one representative coil (the remaining seven coils are symmetrically arranged). Each feeding port is equipped with a matching network that consists of a parallel ($C_p$) and a series ($C_s$) capacitor. $32$ variable capacitors ($C_t$) were symmetrically arranged on the $8$ coils and adjusted to ensure resonance at $297.2$ MHz and decoupling between first order neighbors. The second order neighbors were decoupled using $8$ pairs of perfetly counter-wound inductors (with mutual inductance coefficient equal to $0.9$), each with inductance $L = 22$ nH. The proposed design was inspired by a similar coil \cite{chen2018highly}, which proved useful for phantom EP reconstruction, both in simulation \cite{giannakopoulos2019global} and in experiments \cite{serralles2019noninvasive}, but which was too small to scan a human head. 
\par
We simulated the RF coil array using the surface integral equation (SIE) method, since it can be easily coupled with VIE \cite{villena2016fast}. The current on the surface of the coil was expanded with the RWG basis functions \cite{rao1982electromagnetic}, which have proven to be crucial for SIE accuracy and feasibility. The resulting singular integrals in the Galerkin Method of Moments matrix were calculated semi-analytically with the open source package DIRECTFN \cite{polimeridis2013directfn}. The feeding ports and the lumped elements of the coil were modeled with the delta-gap method \cite{jiao1999fast}. Our implementation of the VIE-SIE method is based on code available in the open-source MARIE software package \cite{villena2015marie}.

\subsection{Tuning-Matching-Decoupling Optimization}

Our approach to properly adjusting the values of the matching and tuning capacitors is similar to the method described in \cite{kozlov2009fast}. First, we assumed all variable capacitors to be short-circuits and the feeding ports to have no matching circuit. As a result, the coupled VIE-SIE solver returned a $40 \times 40$ $Y_n$ admittance parameters matrix, which models the high-frequency electromagnetic phenomena \cite{eleftheriades1996network} and includes the known lumped elements \cite{jiao1999fast}. Second, we treated the $Y_n$ parameter matrix as a black-box $n$-port network containing $m<n$ feeding ports, with indices $p$, and $n-m$ tuning capacitors, with indices $l$:
\begin{equation}
Y_n = \begin{bmatrix}
Y_n^{pp} & Y_n^{pl} \\
Y_n^{lp} & Y_n^{ll} 
\end{bmatrix}.
\end{equation} 
Third, we plugged the variable capacitors into the previously short-circuited ports and attached the matching network to each feeding port. As a result, the $Y_n$ matrix was reduced to a smaller $Y_m$ matrix, from which the scattering parameter matrix $S_m$ can be calculated (see equations below). This reduction was performed multiple times within an iterative optimization that updates the value of the lumped elements to search for the smallest value of the Frobenius norm of the scattering parameter matrix $S_m$. This optimization step simultaneously performed tuning, matching and decoupling at one specified frequency, as described below.
\par    	
$S_m$ is related to the impedance parameter matrix $Z_m$ by
\begin{equation} 
\begin{aligned}
S_m(C_t,C_p,C_s) &= \left(Z_m(C_t,C_p,C_s)+z_0 I\right)^{-1} \\
&\left(Z_m(C_t,C_p,C_s)-z_0 I\right),
\end{aligned}
\end{equation} 
where $z_0$ is the characteristic impedance of the feeding cable, usually set to $50 \Omega$, and $I$ is the identity matrix. $Z_m$ is the inverse of $Y_m$, which can be written as
\begin{equation} 
\begin{aligned}
Z_m(C_t,C_p,C_s) &= Y_m^{-1}(C_t,C_p,C_s) \\ 
&=Y_m^{-1}(C_t,C_p) + \text{diag}\left(\frac{1}{\mathrm{i}\omega C_s}\right),
\end{aligned}
\end{equation} 
where $\omega$ is the angular frequency and $\mathrm{i}$ is the imaginary unit. The entries of $Y_m(C_t,C_p)$ are related to those of $Y_n$ (output of the VIE-SIE solver) by
\begin{equation}
\begin{aligned} 
Y_m(C_t,C_p) &= Y_n^{pp} + \text{diag}\left(\mathrm{i}\omega C_p\right) - Y_n^{pl} Y_L^{-1} Y_n^{lp} \\
Y_L &= \left( Y_n^{ll} + \text{diag}\left(\mathrm{i}\omega C_t\right) \right).
\end{aligned}
\end{equation} 
In the previous equation, the lumped element ports ($l$) were terminated with the tuning capacitors $C_t$ using the Schur complement, and the parallel matching capacitors $C_p$ were included in the feeding ports. These quantities were used to construct the cost function 
\begin{equation} 
f(C) = \frac{1}{2} \norm{W \circ S_m(C)}^2_{\text{fro}},
\end{equation} 
where $W$ is a constant matrix of weights that depends on what values we want to prioritize during the minimization of the scattering parameters. For the presented coil, $W$ was set to a lower triangular matrix of ones, since $S_m$ is symmetric. The gradient of the cost function is shown in the Appendix. 
\par
The ideal capacitor values resulting from the tuning, matching and decoupling optimization were used to terminated the short-circuited ports of the coil. The $8$ different surface currents $\vec{j}^c_p, \: p = 1,\dots,8$ on the coil conductors were individually calculated, using $1$ Voltage signal amplitude at one feeding port at the time, while the rest were terminated with their respective matching circuit in parallel to the impedance of the feeding cable ($50 \Omega$). 

\subsection{RF Shimming}
The incident electric and magnetic field distributions in the space occupied by the head model can be calculated by projecting the surface currents through the $\Nt$ and $\Kt$ operators, respectively:
\begin{equation}
\vec{e}_{\text{inc},p} = \Nt(\vec{j}^c_p), \: \vec{h}_{\text{inc},p} = \Kt(\vec{j}^c_p).
\label{eq:n1}
\end{equation}
From the incident fields, we can compute the magnetic field distribution of each coil (eight separate VIE simulations) inside the head model and generate the corresponding synthetic $B_1^+$ maps that can be used directly (``one at a time'') in GMT. 
\par
Since GMT results depends on the orthogonality of the $B_1^+$ maps, we tried to further improve the conditioning of the problem by performing an ``SVD shimming'' optimization, to calibrate the driving voltages at each coil. The SVD was applied on the reshaped matrix of the $B_1^+$ maps $B^{n \times 8} = USV^*$, where $n$ is the number of voxels, and the resulting left singular vectors $U$ were used as the orthonormalized $B_{1 \: \text{svd}}^+$ maps. To obtain the correct excitation pattern for this shimmed $B_1^+$ set, we multiplied the incident field matrices with a matrix of weights $Q = VS^{-1}$. Note that, while this approach is expected to optimize GMT reconstructions for a given array, if the problem is not well-conditioned to begin with (e.g., in the case of a poorly tuned, matched and decoupled coil array), the optimization could yield elements of the $Q$ matrix significantly different from each other, which would result in large variations between the voltages driving each port. This could then be difficult to achieve in practice, leading to unwanted biases between the experimental $B_1^+$ maps and those synthetically generated by solving the forward problem, which would negatively affect GMT in actual experiments.   
\par
We also tested another RF shimming configuration (``all but one''), which had proven useful in minimizing nulls in $B_1^+$ maps for another EP reconstruction technique \cite{sodickson2012local}. More specifically, we first combined the $B_1^+$ maps from each channel to achieve constructive interference at the center of the object (i.e., by cancelling out all phases at the central voxel), then we created eight new $B_1^+$ maps by subtracting, one at a time, the individual coils' $B_1^+$ from the combined map. We then used the combined map and the first seven ``all but one'' maps as the eight $B_1^+$ maps for the simulated GMT experiment. In an actual GMT experiment, this RF shimming approach could be achieved by calculating the RF shims to constructively combine all $B_1^+$ at the center and turning off (i.e., setting voltage to zero for) one channel at a time to create the distinct excitations. 

\section{Numerical Results} \label{sc:IV}

\subsection{Stability of Tuning-Matching-Decoupling}

The $S_m$ parameter matrix, after loading the triangular coil with the ``Duke'' head model and discretizing the conductors with $2026$ triangular elements, is shown in Fig. 2. The small values in the diagonal elements ($-32.59$ to $-17.36$ dB), indicate good tuning and matching at $297.2$ MHz, whereas the small off-diagonal elements ($-25.86$ to $-13.31$ dB) confirm that the channels were well-decoupled. The values of the tuning capacitors were between $1$ and $5$ pF, which is a typical range for $7$ T coils.  

\renewcommand{\thefigure}{2}
\begin{figure}[ht!]
\begin{center}
\includegraphics[width=0.48\textwidth, trim={0 0 0 0}]{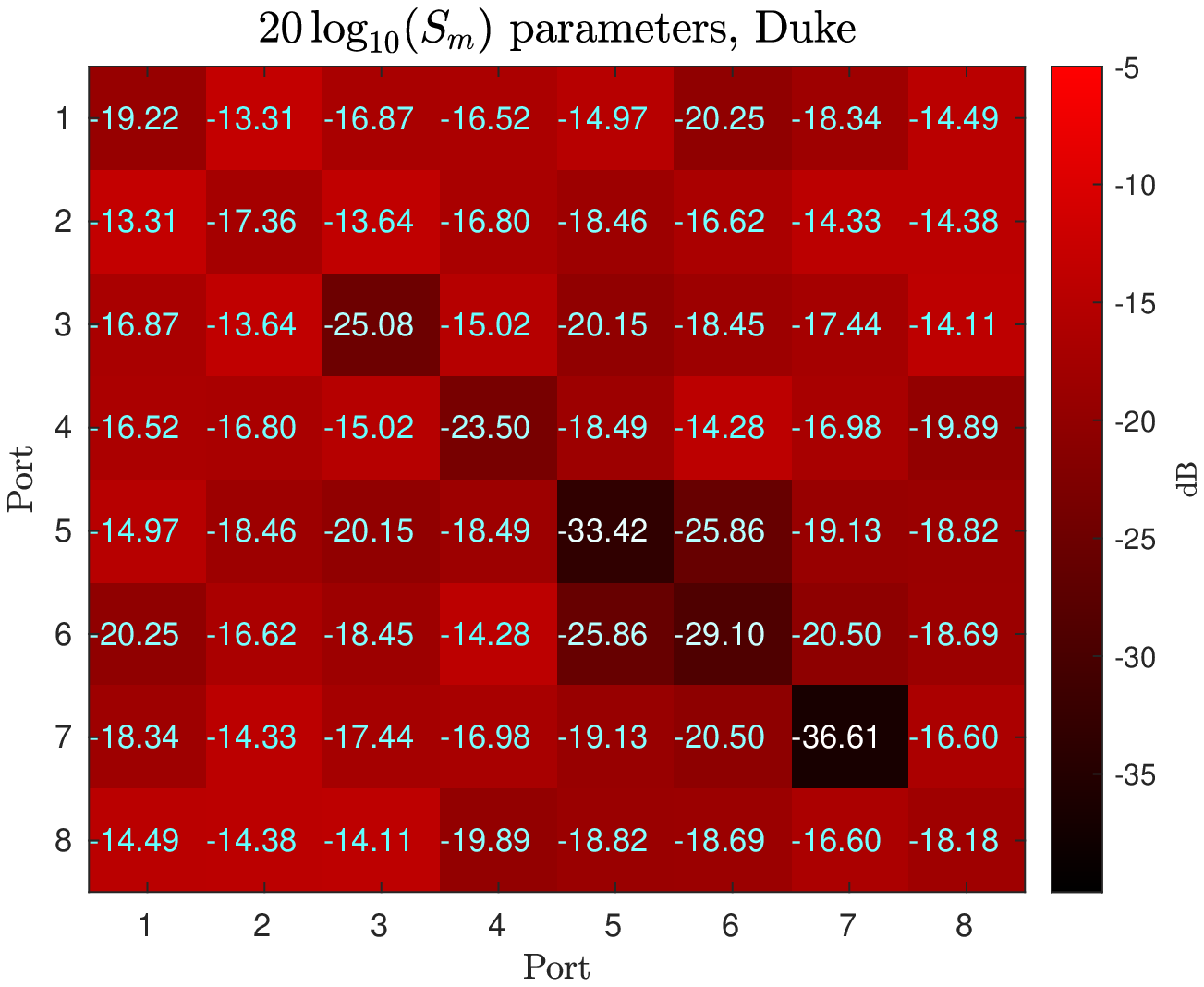}
\caption{The $S_m$ parameters of the RF coil array loaded with the "Duke" head model, after the optimization.}\label{fig:n2}
\end{center}
\end{figure}

Since geometric decoupling \cite{roemer1990nmr}, which is broadband, was not used, we assessed the sensitivity of the $S_m$ parameters with respect to the load. To do so, we calculated the $S_m$ after loading the same coil with a smaller head model (``Ella'', from the Virtual family population \cite{VirtualFamily}), which had a different shape and interior structure than the head model used to tune and match the coil. In particular, ``Ella'' contains 20 tissues with unique EP and it was enclosed by a domain with dimensions equal to $17.5 \times 21 \times 24$ cm$^3$, whereas ``Duke'' has 21 tissues and its domain was $19 \times 23.5 \times 23$ cm$^3$. The new setup and the corresponding $S_m$ are shown in Fig. 3. Tuning and matching remained overall stable, whereas the value of the off-diagonal elements $S_{12}, S_{18}$ and $S_{23}$ slightly increased to approximately $-10$ dB. This was expected, since first-order neighbor decoupling is accomplished through tuning capacitors and, therefore, is affected by the loading condition.

\renewcommand{\thefigure}{3}
\begin{figure}[ht!]
\begin{center}
\includegraphics[width=0.48\textwidth, trim={0 0 0 0}]{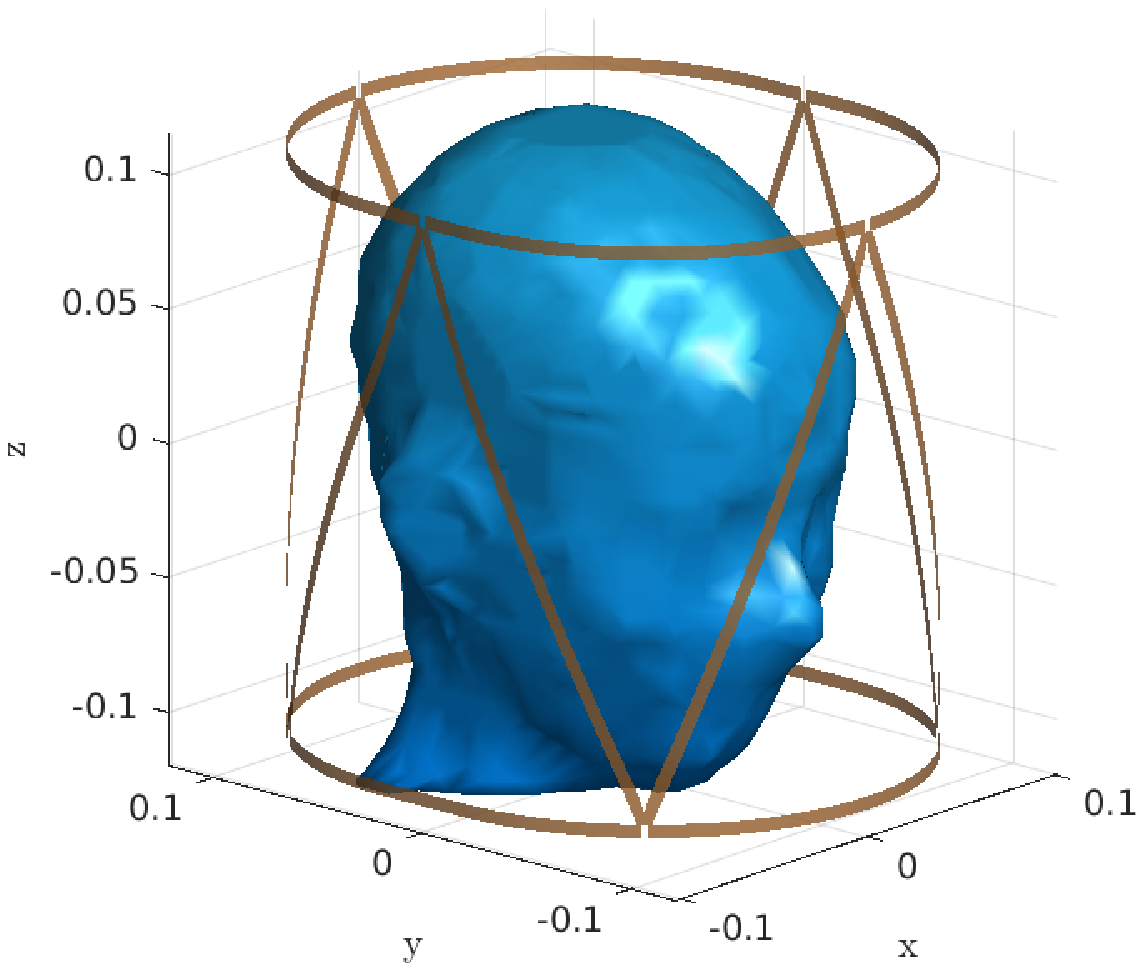}\\
\includegraphics[width=0.48\textwidth, trim={0 0 0 0}]{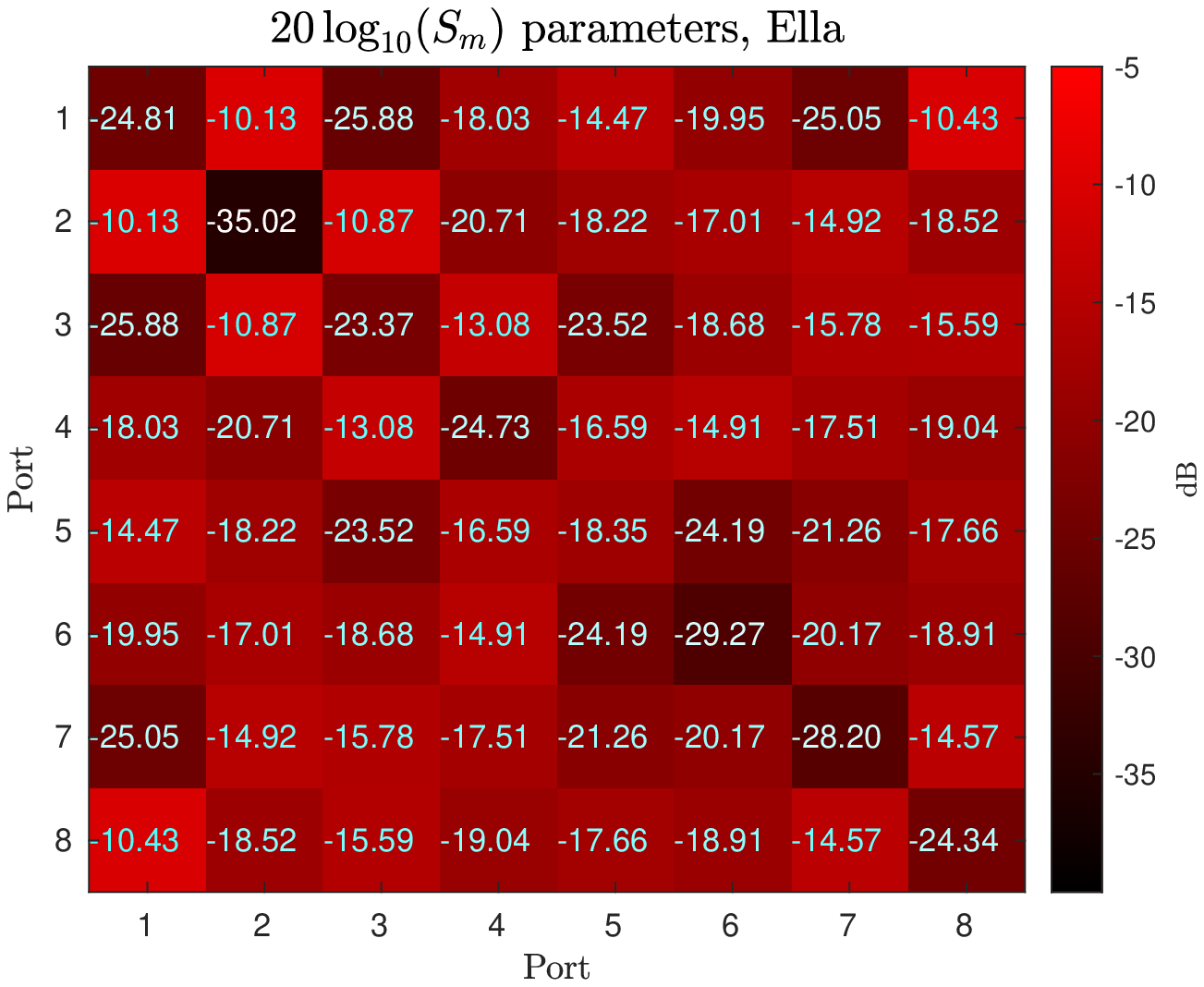}
\caption{The RF coil array loaded with "Ella" head model (top), and the corresponding $S_m$ parameters (bottom). The coil was tuned using the ``Duke'' model.}
\label{fig:n3}
\end{center}
\end{figure}

\subsection{Electrical Property Reconstruction}

Figs. 4 and 5 show the absolute values and the relative phases of the $B_1^+$ of each channel for an axial slice through the ``Duke'' head. The condition number of the matrix $B^{n \times 8}$, obtained by reshaping the $B_1^+$ maps for all coils and all $n$ voxels, was $\sim 4.75$. The condition number was instead $1$ and $40$ for the ``SVD shimming'' and the ``all but one'' calibrations (see III. B), respectively. As in \cite{serralles2019noninvasive}, we corrupted the $B_1^+$ maps with white Gaussian noise with peak SNR of $200$, corresponding to mean SNRs ranging between $82$ and $88$ for the $8$ channels. 

\renewcommand{\thefigure}{4}
\begin{figure}[ht!]
\begin{center}
\includegraphics[width=0.48\textwidth]{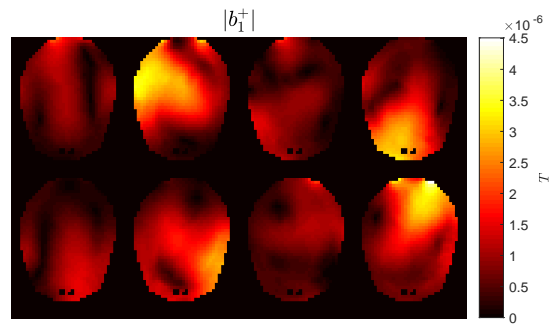}
\caption{$\abs{B_1^+}$ of each coil array element using one port at a time for excitation for an axial section of the ``Duke'' head model. The field is masked outside the phantom for enhanced visualization.}\label{fig:n4}
\end{center}
\end{figure}

\renewcommand{\thefigure}{5}
\begin{figure}[ht!]
\begin{center}
\includegraphics[width=0.48\textwidth]{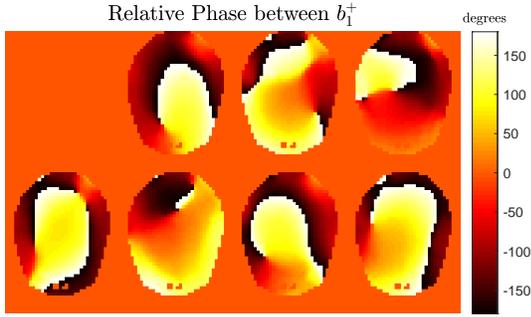}
\caption{Phase of the $B_1^+$ field of each coil, relative to the phase of the $B_1^+$ field of the first coil, for an axial section of the ``Duke'' head model. Values are masked outside the phantom for enhanced visualization.}\label{fig:n5}
\end{center}
\end{figure}

GMT was executed for $500$ iterations, starting from an initial guess of spatially homogeneous EP. Fig. 6. shows the reconstructed EP for the three RF shims. While the internal structure of the head was preserved, some degree of blurring was observed between specific regions, e.g., at grey and white matter boundaries. This is due to non-optimal settings of Match Regularization, which were copied from previous work \cite{serralles2019noninvasive} that employed EM basis functions rather than realistic coils for the MR excitation. For the ``one at a time'' approach, we performed the GMT optimization also for ``Ella'', both without re-tuning the coil and after re-tuning it (Fig. 7). Both reconstructions resulted in similar EP maps, which confirms the stability of the coil tuning when loaded with a smaller head (see Fig. 3). Table I reports the mean of the peak-normalized absolute error (PNAE, i.e., the difference between the ground truth and reconstructed EP, divided by the maximum value of the ground truth EP), calculated over all voxels of the head. Note that the overall error after re-tuning the coil for ``Ella'' only slightly improved. 

\begin{table}[!ht]
\caption{EP reconstruction mean PNAE} \label{tb:n1} \centering
{\def\arraystretch{1.5}\tabcolsep=3pt

\begin{tabular}{ c|c|c|c|c }
\hline
\hline
Excitation Pattern                  & EP           & ``Duke''  &  ``Ella'' w/ retuning &  ``Ella'' w/o retuning \\
\hline
\multirow{2}{*}{SVD calibration}    & $\epsilon_r$ & $5.7 \%$  & NA                    & NA \\     
						            & $\sigma_e$   & $3.6  \%$ & NA                    & NA \\     
\hline
\multirow{2}{*}{One at a time}      & $\epsilon_r$ & $7.5 \%$  & $9.1 \%$              & $9.5 \%$ \\     
	           					    & $\sigma_e$   & $4.8  \%$ & $7.1 \%$              & $7.5 \%$\\     
\hline
\multirow{2}{*}{All but one}        & $\epsilon_r$ & $11.4 \%$ & NA                    & NA \\     
	           					    & $\sigma_e$   & $7.0 \%$  & NA                    & NA \\     
\hline
\hline
\end{tabular}
}
\end{table}       

\renewcommand{\thefigure}{6}
\begin{figure}[ht!]
\includegraphics[width=0.47\textwidth]{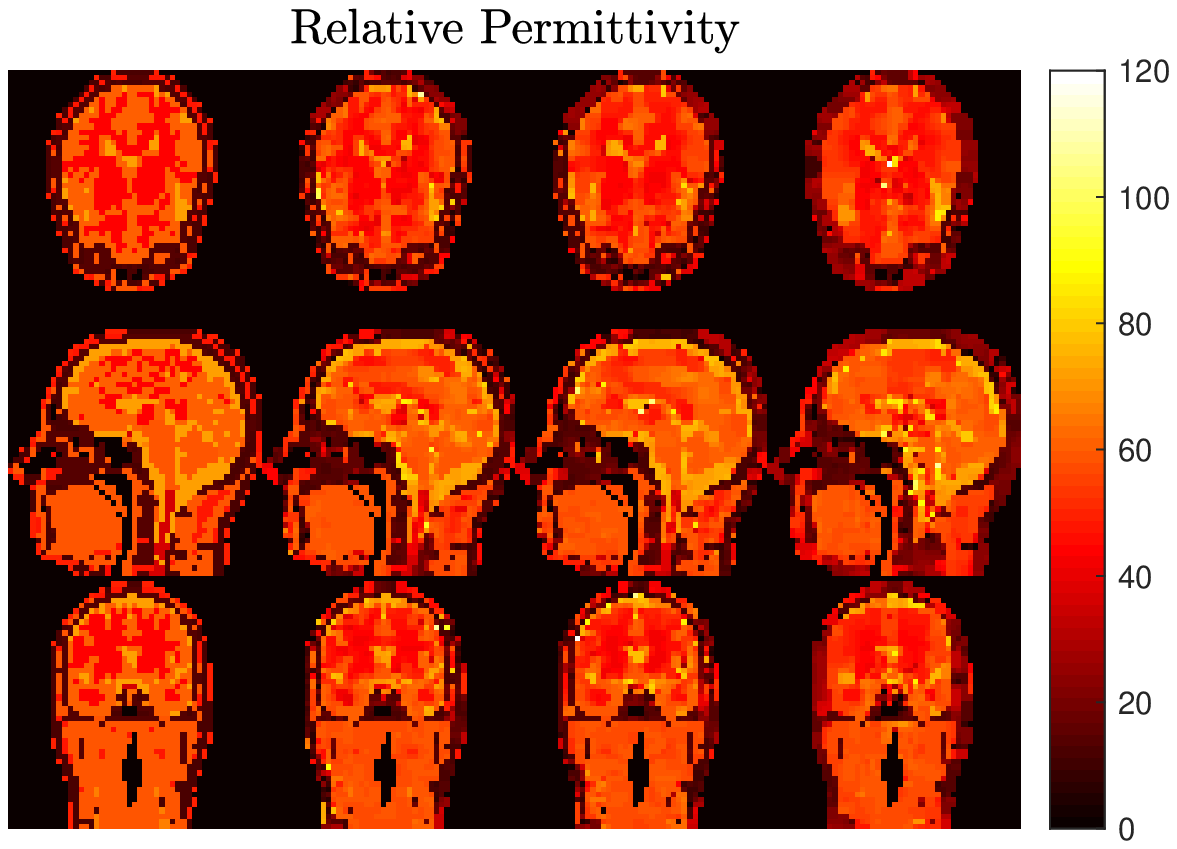} \\[-1cm]
\includegraphics[width=0.47\textwidth]{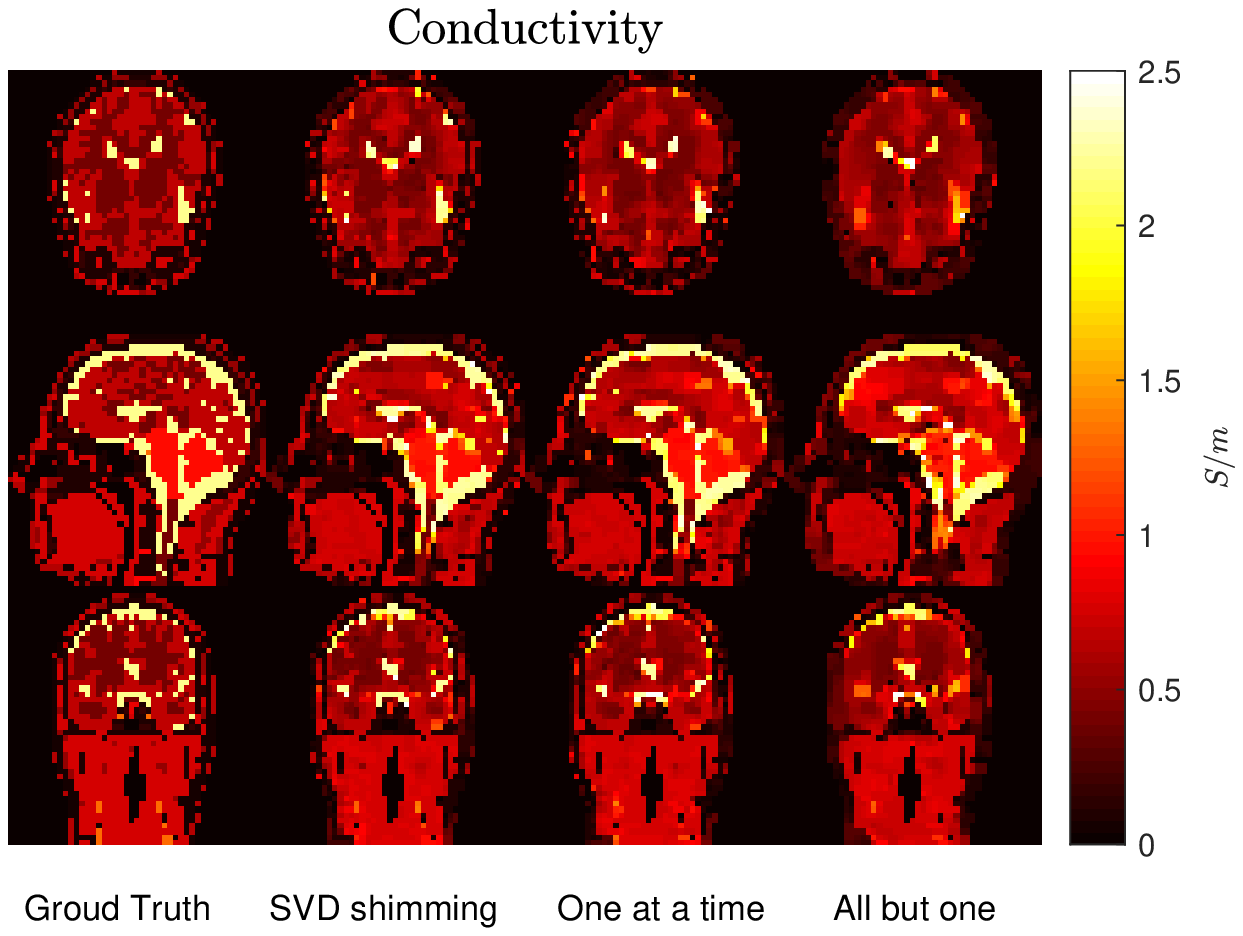}
\caption{Reconstructed relative permittivity (top) and conductivity (bottom) maps for an axial, a sagittal and a coronal slice through the "Duke" head model, using three different excitation patterns. From left to right, ground truth EP are compared with results for ``SVD shimming'', ``one at a time'', and ``all but one'' excitations.}\label{fig:n6}
\end{figure}

\renewcommand{\thefigure}{7}
\begin{figure}[ht!]
\includegraphics[width=0.47\textwidth]{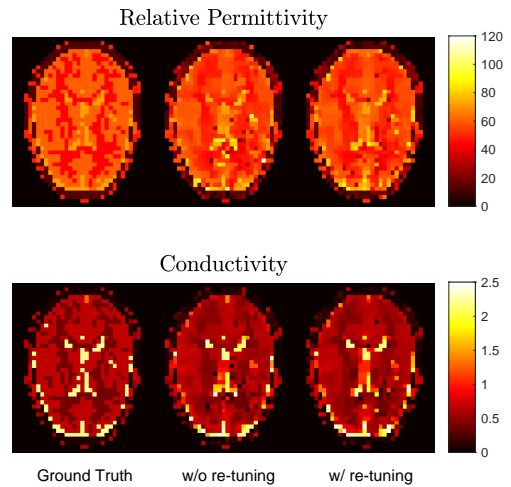} 
\caption{Relative permittivity (top) and conductivity (bottom), reconstructed both without re-tuning and after re-tuning the coil, are compared with corresponding ground truth EP, for an axial plane of ``Ella'', using the ``one at a time'' excitation pattern.}\label{fig:n7}
\end{figure}

Fig. 8. shows histograms of the PNAE for all voxels in the ``Duke'' head model, for the three excitation patterns. As for previous work \cite{serralles2019noninvasive}, PNAE was preferred over relative error, to avoid the histogram being biased by high relative errors at voxels where EP values are small. 

\renewcommand{\thefigure}{8}
\begin{figure}[ht!]
\includegraphics[width=0.48\textwidth]{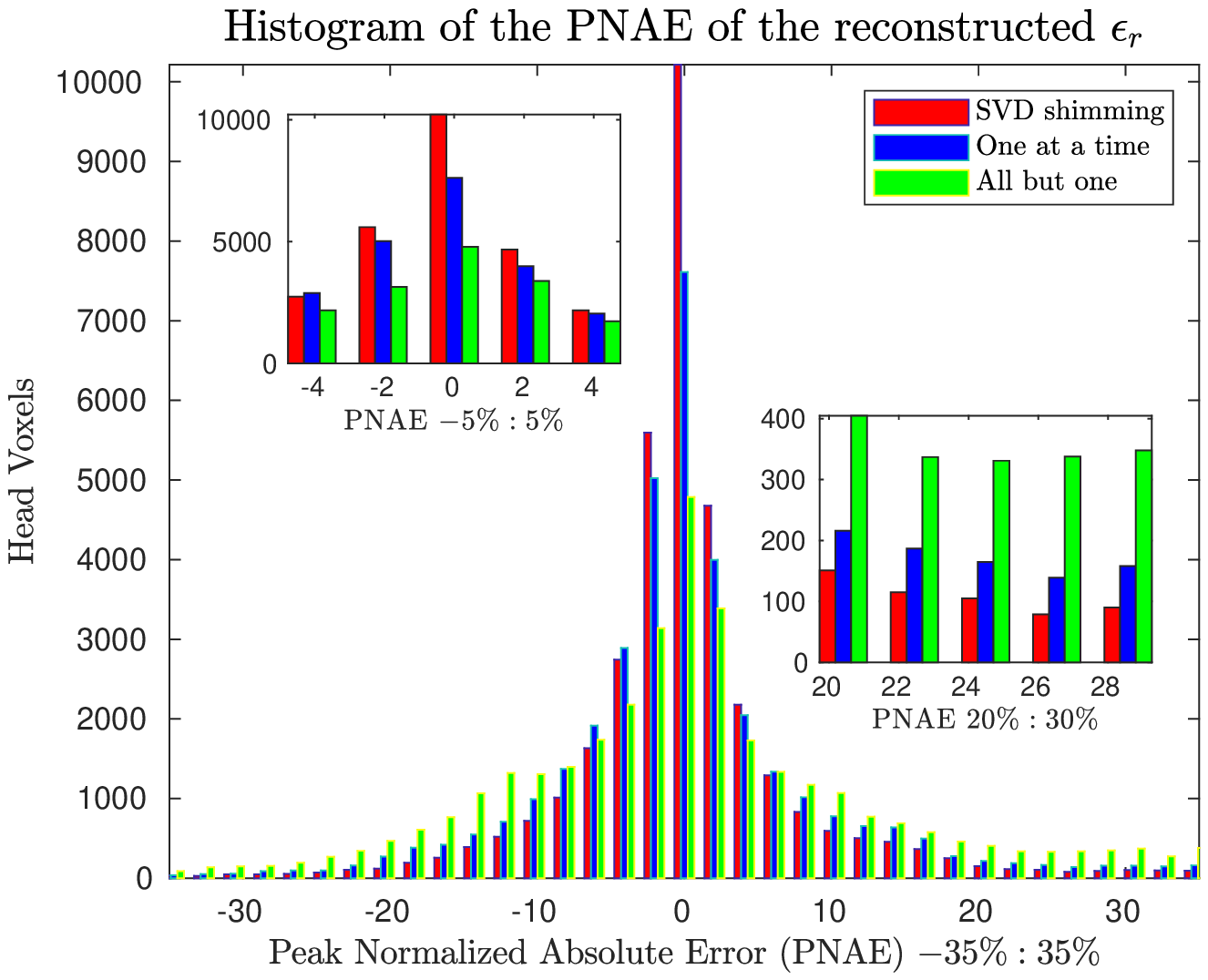} \\
\includegraphics[width=0.48\textwidth]{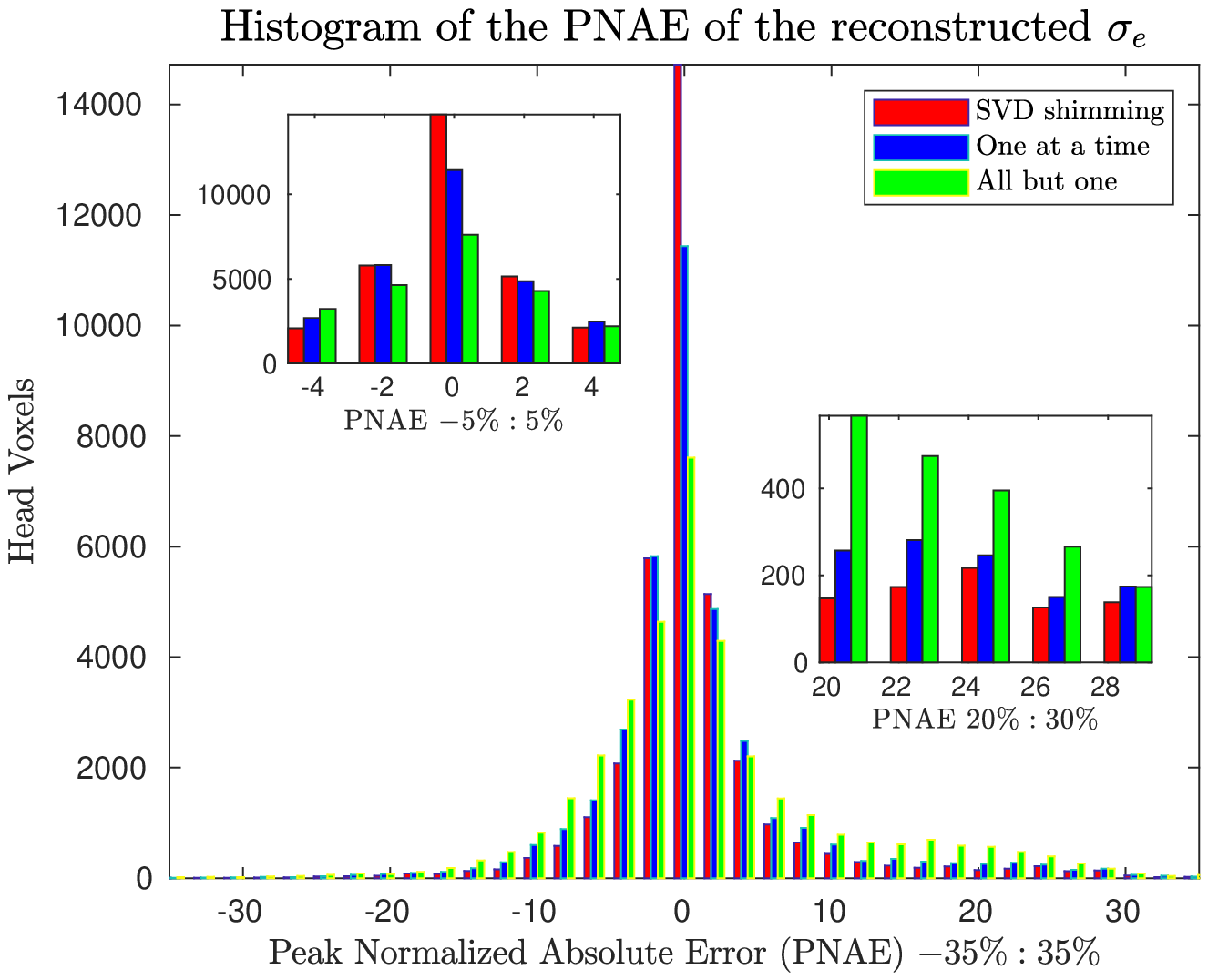}
\caption{Histograms of the peak normalized absolute error for all voxels in the ``Duke'' head model. The error distribution is shown for relative permittivity (top) and conductivity (bottom). Zoomed view of two regions of the histograms are also shown to highlight differences among the three excitation patterns.}\label{fig:n8}
\end{figure}

\subsection{Estimated EM Field Distribution}

We used the JVIE solver to assess how the error in the reconstructed EP propagates in the estimation of EM fields distributions. Specifically, we shimmed the coils' incident fields to illuminate the head with a total incident field that resembled the circularly polarized mode of a birdcage coil \cite{hayes1985efficient}. Using this excitation and the EP reconstructed with the ``one at a time'' approach, we computed the $\abs{B_1^+}$ distribution, as well as the absorbed power at each voxel $p_{\text{abs}} = \frac{1}{2} \sigma_e \abs{\vec{e}}^2$, where $\vec{e}$ is the electric field. Fig. 9 compares the resulting maps with those obtained using the ground truth EP.

\renewcommand{\thefigure}{9}
\begin{figure}[ht!]
\includegraphics[width=0.48\textwidth]{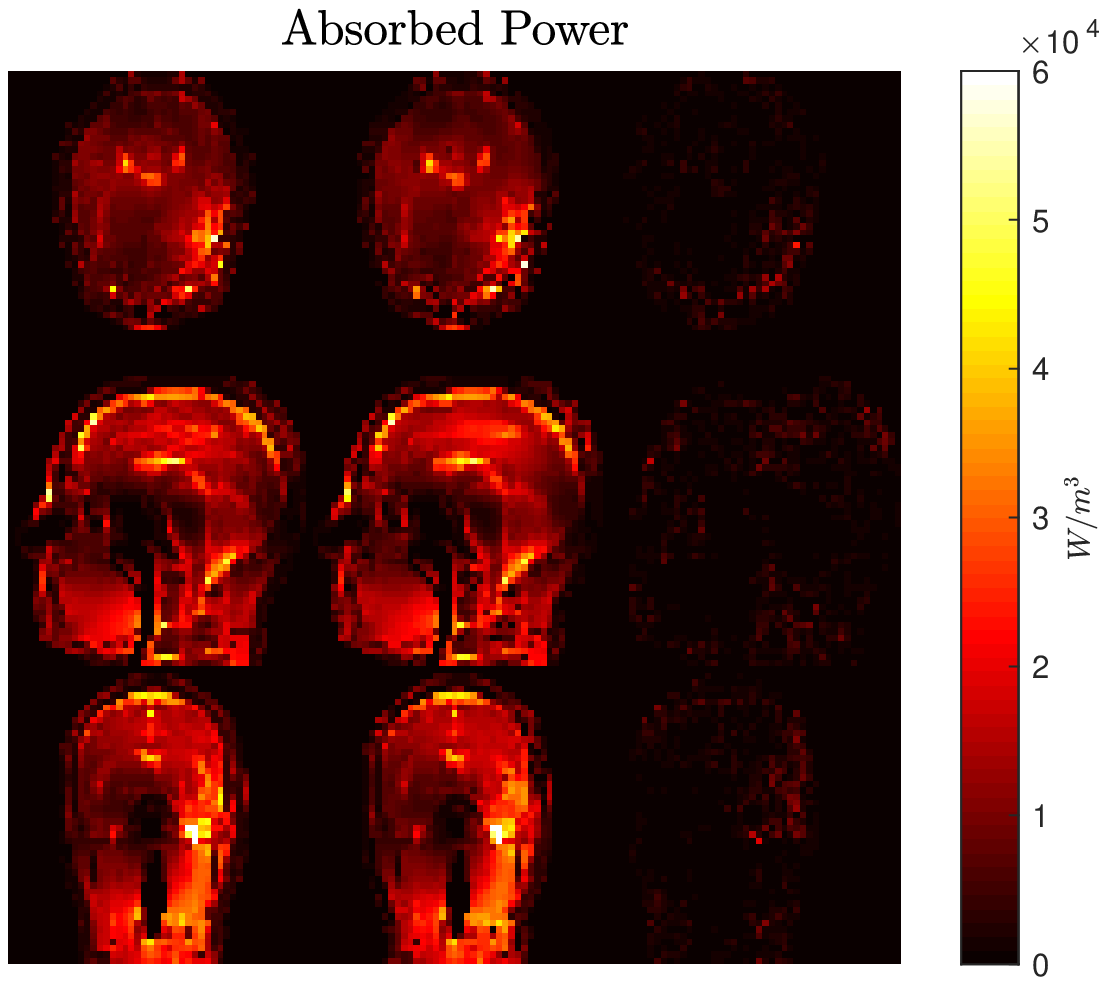}\\
\includegraphics[width=0.48\textwidth]{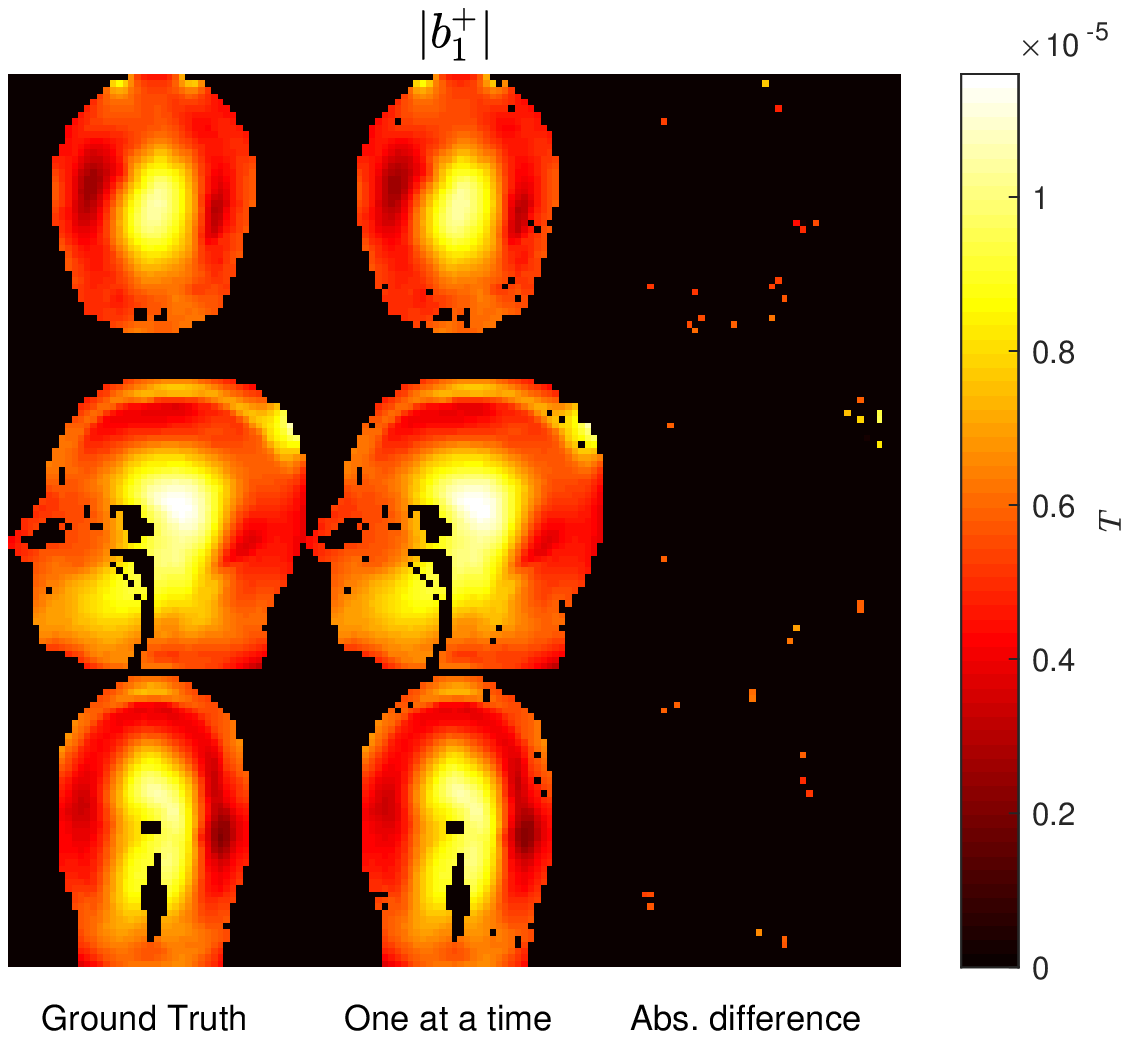}
\caption{Absorbed power (top) and $\abs{B_1^+}$ (bottom) for the central axial, sagittal and coronal planes of the ``Duke'' head model. Maps were calculated for a circularly polarized excitation using the ground truth EP (left) and the EP reconstructed with the ``one at a time'' excitation pattern (middle). A map of the absolute difference between the first two columns is shown for each estimated measurement (right).}\label{fig:n9}
\end{figure}

The PNAE histograms in Fig. 10. shows that the error for the absorbed power was mostly within $-5\%$ and $5\%$, whereas for the $\abs{B_1^+}$ the error was strictly between $-1\%$ and $1\%$. Table II provides an overall quantitative comparison, including the number of iterations required by the JVIE solver to converge in each case, as well as the scalar absorbed and scattered power, calculated with the formulas in \cite{polimeridis2014computation}.  

\renewcommand{\thefigure}{10}
\begin{figure}[ht!]
\includegraphics[width=0.48\textwidth, trim={0 0 0 0}]{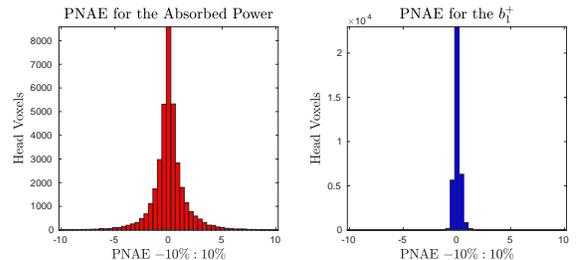}
\caption{Histogram of the peak normalized absolute error for the absorbed power (left) and the $\abs{B_1^+}$ (right) calculated using the electrical properties reconstructed with the ``one at a time'' approach.}\label{fig:n10}
\end{figure}

\begin{table}[!ht]
\caption{Comparison of EM field related estimated quantities} \label{tb:n2} \centering
{\def\arraystretch{2}\tabcolsep=3pt

\begin{tabular}{ c|c|c }
\hline
\hline
Measurement            & Ground Truth EP  & Reconstructed EP  \\
\hline
mean PNAE $p_{\text{abs}}$ & NA               & $0.46 \%$         \\
mean PNAE $\abs{B_1^+}$    & NA               & $0.38 \%$         \\
\hline
Absorbed Power (Watt)  & $52.81$          & $52.54$           \\
Scattered Power (Watt) & $6.56$           & $6.48$            \\
\hline
Iteration Count        & $183$            & $266$             \\     
\hline
\hline
\end{tabular}
}
\end{table}

\subsection{Brain Tumor Detection}

It is has been shown that the conductivity of cancerous tissue differs from that of neighboring healthy tissue\cite{o2007dielectric, surowiec1988dielectric} and can be three times higher than that of white matter in the case of brain tumors \cite{huhndorf2013systematic}. To evaluate the diagnostic performance of GMT in a simulated clinical scenario, we inserted a numerical tumor into the ``Duke'' head model. Based on \cite{yoo2004dielectric}, we used relative permittivity and conductivity of $68$ and $1.1$ S/m, respectively. We performed two numerical GMT experiments using peak SNR of $200$ and stopping after $500$ iterations. In the first experiment (Case I), we re-tuned the coil after inserting the synthetic tumor, used the ``SVD shimming'' excitation pattern and the ground truth EP (without the tumor) as the initial guess, to mimic the detection of a tumor in an otherwise healthy brain. In the second experiment (Case II), we did not re-tune the coil, used the ``one at a time'' excitation pattern, and we started from a homogeneous initial guess equal to the average EP among all voxels. Case I was included to assess the sensitivity of GMT for the detection of a realistic tumor, whereas Case II provides a worst-case scenario that may present in future in-vivo applications. The reconstructed EP maps are compared with the true EP distribution in Fig. 11, showing that the tumor was correctly detected in both cases. In Case II, the reconstruction is slightly blurrier.

\renewcommand{\thefigure}{11}
\begin{figure}[ht!]
\includegraphics[width=0.47\textwidth, trim={0 0 0 0}]{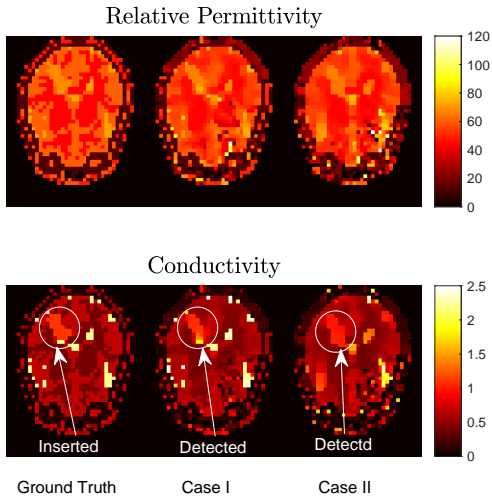}
\caption{Relative permittivity (top) and conductivity (bottom) maps for an axial slice through the ``Duke'' head model with a numerically inserted tumor. Ground Truth EP (left) are compared with GMT reconstructions for Case I (middle) and Case II (right).}\label{fig:n11}
\end{figure}

Fig. 12 (top) compares the original tumor with the tumor segmented from the conductivity map reconstructed with GMT (values between $0.8$ and $1.2$). The histogram in Fig. 12 (bottom) shows the values of the conductivity for the segmented tumor voxels. The tumor was accurately detected in both cases, although outliers in the histograms suggest that its boundaries were blurred by the GMT reconstruction, due to sub-optimal noise regularization. Note that the histogram of the estimated conductivity of the tumor did not cluster around its true value ($1.1$ S/m) for Case II, but around $1.0$ S/m, which is still above the average brain conductivity and, therefore, would not impair cancer detection.

\renewcommand{\thefigure}{12}
\begin{figure}[ht!]
\includegraphics[width=0.47\textwidth, trim={0 0 0 0}]{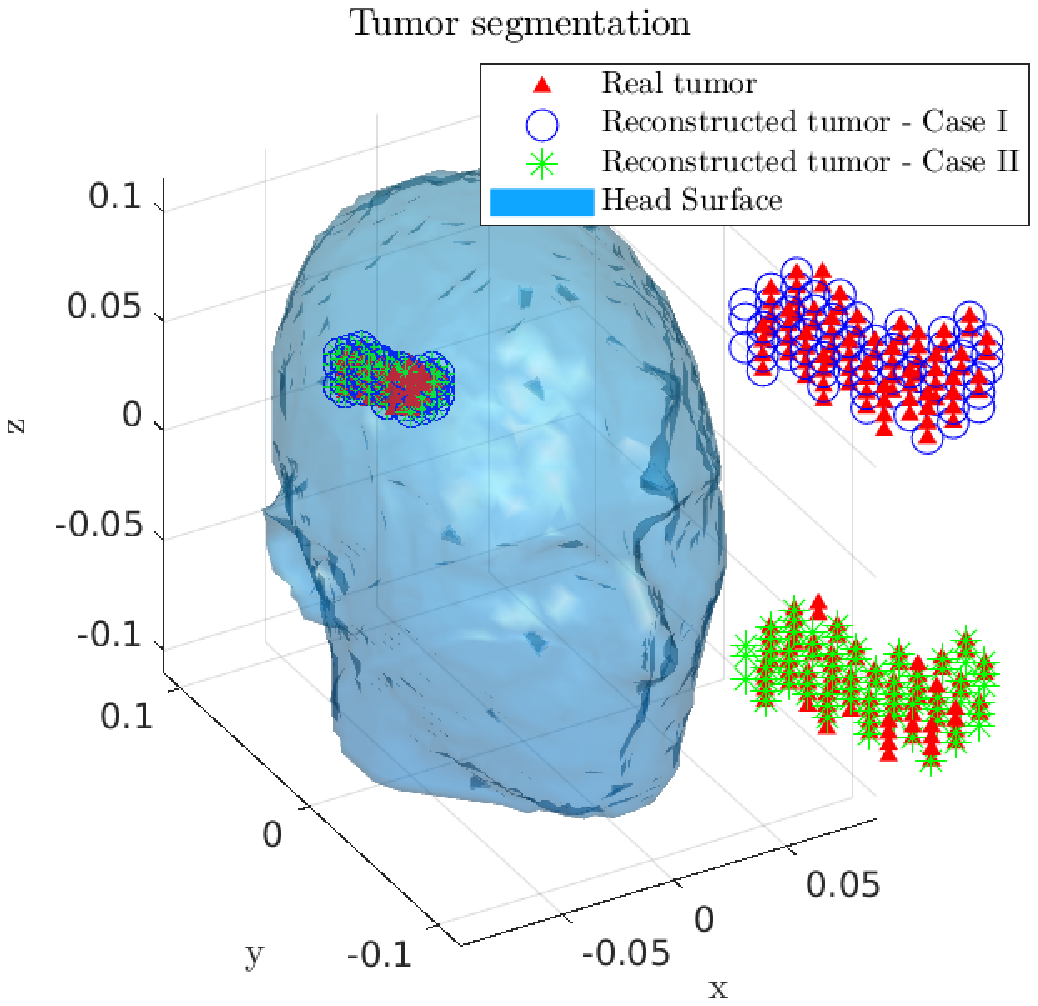}  \\
\includegraphics[width=0.47\textwidth, trim={0 0 0 0}]{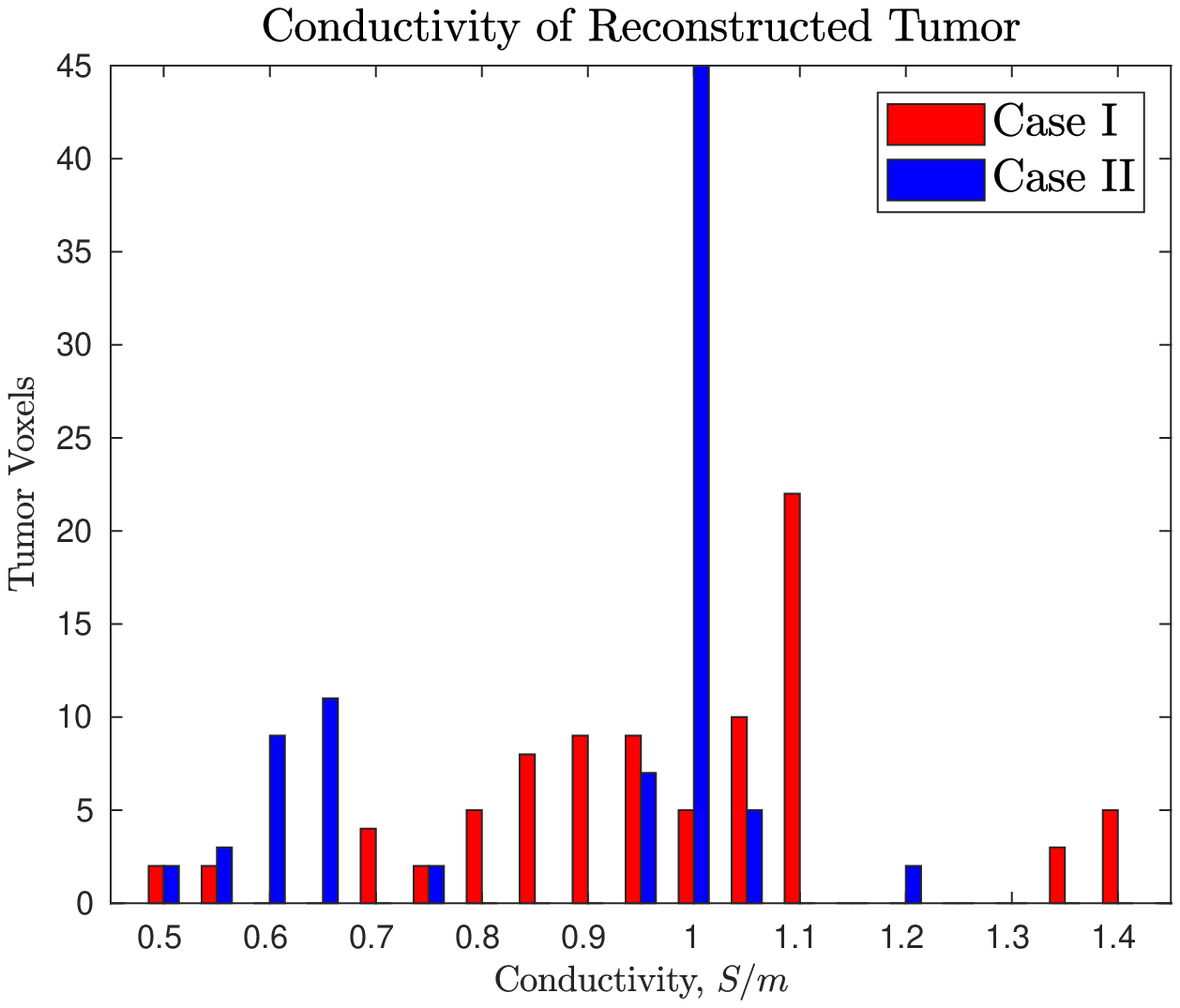} 
\caption{GMT results for a simulated brain tumor. (top) The numerically inserted tumor is compared with the tumor segmented from the conductivity maps reconstructed in Case I and Case II. (bottom) The histogram of the reconstructed conductivity for the tumor voxels was clustered around the true value of $1.1$ S/m for Case I, and $1.0$ for Case II.}\label{fig:n12}
\end{figure}

\section{Discussion} \label{sc:V} 
GMT is a recently introduced volumetric technique for the estimation of electrical properties from MR measurements. Results of phantom experiments and simulations with heterogeneous anatomical models have been promising. However, until now numerical tests had been performed using ideal incident fields obtained from an EM basis. Therefore, the feasibility of performing GMT in-vivo has not yet been thoroughly assessed. The aim of this work was to design an 8-channel transmit-receive RF coil and use it to evaluate GMT in simulation for realistic in-vivo scenarios.
\par
 Furthermore, this work showed that for an RF array to work well with GMT, it needs to generate distinct transmit fields, which reflect the inhomogeneity of the underlying EP distribution while providing acceptable SNR throughout the volume of the object of interest. Note that this requirement pulls in the opposite direction of traditional coil design, which aims at maximizing transmit field homogeneity. We fulfilled the above requirements by arranging $8$ triangularly shaped coils symmetrically around a quasi-elliptical cylindrical surface. One feature of the proposed triangular coil design is a non-trivial $B_z$ component. While having a pronounced $B_z$ gradient that depends on the underlying EP distribution improves the numerical conditioning of GMT, note that it is highly undesired for various other EP mapping techniques that make symmetry assumptions on the distribution of the EM field \cite{katscher2009determination,mandija2018error}. We proposed a metric to evaluate the array's suitability for GMT by looking at the condition number of a matrix containing the transmit fields at every voxel to check their distinctness. The condition number for the triangular array was approximately four, which is not too far from the ideal case of the basis elements (condition number = $1$). As a result, the mean PNAE over the ``Duke'' head was $7.5\%$ and $4.8\%$ after 500 GMT iterations exciting with one coil element at a time, compared with $3.82\%$ and $2.33\%$ when exciting with an ultimate EM basis a smaller head model and letting GMT run for $1000$ iterations \cite{serralles2019noninvasive}, for relative permittivity and conductivity, respectively.
\par
We also showed that the condition number could be manipulated by RF shimming. In particular, an ``all but one'' RF shimming, which was found to be useful for differential EP reconstruction techniques \cite{sodickson2012local}, worsens the condition number by a factor of ten, yielding less accurate results for GMT (see Table I). Even in the central voxels, where this RF shimming creates a large SNR, the reconstruction failed (see Fig. 6), because the associated phase cancellation made the inverse problem more ill-posed due to the lack of inhomogeneity in the $B_1^+$. On the other hand, an SVD-based shimming could bring the condition number to one, although the effect on the accuracy of GMT was minimal, suggesting the near optimality of the designed coil. Furthermore, we must note that it could be difficult to apply the ``SVD shimming'' in actual experiments, since it would require a lengthy pre-scan to obtain the $B_1^+$ of each coil and the SVD could result in a wide range of values for the amplitude and phase RF shims at each port, which may be difficult to achieve in practice. An alternative approach to further reduce the condition number could be improving decoupling between array elements. This could be achieved by including the inductor pairs and the mutual inductance coefficient in the tuning-matching-decoupling optimization (Section III.A), or by using a more complex circuit at each port \cite{mahmood2016general}. However, both solutions could make it more difficult to match simulated and experimental incident fields, which is critical for GMT accuracy \cite{serralles2019noninvasive}. In addition, it is not clear by how much the EP estimation would improve, since the errors we observed in this work could be due to non optimal settings of the regularizer, rather than imperfect excitations. 
\par
As for all inverse problems affected by noise, the performance of the regularizer is important for GMT. Match Regularization performs intrinsically well at boundaries with high contrast where it behaves as an $L_0$ regularizer, but in regions of low contrast, where it behaves as total variation, it can result in blurring, as shown in Fig. 6. This could be improved by manually tweaking the three parameters upon which Match Regularization depends, but the computational cost of running GMT multiple times for different settings have prevented such optimization. In fact, in this work we decided to use the same values for the Match Regularization parameters that were used when the incident fields were generated with an EM basis \cite{serralles2019noninvasive}. A possible alternative that could be explored in future work would involve training a neural network to optimize the regularization term, as was proposed for a different MR-related inverse problem \cite{hammernik2018learning}.
\par
Another approach to improve the conditioning of the GMT inverse problem would be to modify the cost function to include receive sensitivity information as well as transmit fields.  As was demonstrated with the Local Maxwell Tomography technique \cite{sodickson2012local}, the combination of transmit and receive field information is sufficient to resolve fundamental indeterminacies related to absolute field phase. Though GMT uses a specific coil model to circumvent the problem of unknown coil phases, the incorporation of additional receive-related information present in MR signals would likely improve the performance of GMT as well. However, the computation time would considerably increase. 
\par
A slight degree of blurring in regions with small differences in EP may not be crucial if one is interested in using the estimated EP to predict RF field interactions with tissues. In fact, the spatial distribution of the EM field is smooth and may not be affected by small voxel-by-voxel changes in EP. For example, our results (Fig. 9) show that despite an almost $6\%$ error in EP estimation, the associated $\abs{B_1^+}$ and absorbed power distributions could be predicted with average errors smaller than $0.5\%$. Blurring in the estimated EP maps could instead have a larger effect for the diagnosis of focal lesions. However, we showed that GMT could detect an arbitrarily shaped brain tumor, despite sub-optimal regularization. Specifically, for most voxels within the cancer region, the estimated conductivity was sufficiently close to the true $1.1$ S/m value to correctly discriminate the lesion from the surrounding tissue. The largest estimation errors were found at the edges of the tumor, but we expect this would improve by optimizing Match Regularization. Such improvement would be even more critical for the goal of characterizing heterogeneous tumor lesions, rather than simply detecting them.
\par
GMT could be considerably accelerated using a graphics processing unit (GPU), since the most expensive operations are element-wise matrix products and fast Fourier transforms. At the same time, the memory required to store the $\Nt$ and $\Kt$ operators increases with the number of voxels and could rapidly become prohibitive even for modern GPU's. To account for this trade-off and enable rapid execution via GPU, all simulations in this work were performed with a coarse voxel isotropic resolution of $5$ mm. In particular, we used a single Nvidia Titan V GPU card, where each forward solution required approximately $10$ minutes for convergence and the entire reconstruction process required $\sim 6$ days for $500$ GMT iterations. Note that the current implementation of GMT runs on Matlab (version R2019a was used for this work), except for the matrix-vector product, which was implemented in CUDA 10.0. Since it was shown that the performance of GMT is not affected by the size of the voxel \cite{serralles2019noninvasive}, it will be desirable to use clinical resolutions ($\sim 1$ mm) for future in-vivo translation. One possible solution that will be explored to make that feasible is to compress the VIE operators using the Tucker decomposition \cite{polimeridis2014compression, giannakopoulos20183d}, which should allow us to perform the necessary matrix-vector products in GPUs even at high spatial resolutions \cite{giannakopoulos2018memory}. Moreover, GMT could be further accelerated to facilitate in-vivo applications by applying a recently proposed preconditioner for the solution of the forward problem \cite{georgakis2019fast}.
\par
In this work, we used the true EP of the head models to calibrate the current distribution on the coil conductors, which is used to compute the incident fields (see Eq. \ref{eq:n1}). This step consists of a surface-volume integral equation (VSIE) coupling, which was performed before the GMT optimization. Regrettably, this calibration cannot be applied to in-vivo experiments, since the EP of the object are unknown. Alternatively, an initial calibration could be performed with a phantom of known EP, as in \cite{serralles2019noninvasive}, and used for in-vivo GMT experiments. However, since our proposed coil is close fitting, such an approach could lead to erroneous results, because the current distribution would be sensitive to the different head sizes of the subjects. The most accurate approach would be to update not only the guess of EP, but also the currents on the coil conductors at every iteration of GMT. That would require performing an VSIE coupling calculation for each GMT iteration, which with the existing EM solver would be computationally expensive, since the coupling matrix between the coil and the body requires a large amount of memory \cite{villena2016fast}. A novel approach that could avoid this costly step has been recently proposed \cite{guryev2019fast}, in which the triangular mesh of the coil is projected onto voxels, in order to enable using FFT to accelerate the matrix-vector products. This method would remove the computational burden of performing the VSIE coupling step and we plan to incorporate it in GMT before performing in-vivo experiments.
\par
GMT uses a quasi-newton optimization technique, therefore it could be further accelerated by optimizing the choice of the initial guess. For in-vivo applications, one approach could be to segment the main tissues on a suitable MRI acquisition, and pre-assign EP based on literature values \cite{VirtualFamily, gabriel1996dielectricii}. An alternative approach could be based on machine learning. Specifically, one could train a neural network, as, for example, in \cite{mandija2019opening}, with $B_1^+$ as the input and EP as the output. The network would then be able to rapidly (in seconds) generate an initial guess for GMT, considerably reducing the number of iterations.

\section{Conclusion} \label{sc:VI} 
In conclusion, we showed that GMT can accurately estimate EP at $7$ T and detect a simulated brain tumor using an $8$-channel transmit array yielding inhomogeneous and distinct $B_1^+$ distributions. The proposed triangular array could be used for GMT directly, without RF shimming, by exciting one port at time with $1$ Volt. RF power deposition and transmit field distribution in tissues could be accurately predicted despite estimation errors in EP. Future work will focus on building the designed array, matching its $S_m$ parameter matrix with the simulation, and, finally, performing GMT experiments on human subjects.

\renewcommand\appendixname{\textsc{\normalsize Appendix: Gradient of the cost function}}
\appendix[]

The gradient of the cost function ($f$) with respect to the capacitance $c \in \{C_t,C_p,C_s\}$ was obtained via matrix calculus. The partial derivative of $f$ is
\begin{equation}
\begin{aligned} 
\frac{\partial f}{\partial c} &= \frac{1}{2} \text{trace} \Bigl( \Bigl(W \circ \frac{\partial S_m}{\partial c} \Bigr)^{*} \Bigl(W \circ S_m\Bigr) \\
&+ \Bigl(W \circ S_m\Bigr)^{*} \Bigl(W \circ \frac{\partial S_m}{\partial c}\Bigr) \Bigr).
\end{aligned}
\end{equation} 
The partial derivative of $S_m$ over $c$ in this expression can be calculated as
\begin{equation} 
\begin{aligned}
\frac{\partial S_m}{\partial c} &= -\left( Z_m + z_0 I \right)^{-1} \frac{\partial Z_m}{\partial c} S_m \\ 
& + \left( Z_m + z_0 I \right)^{-1} \frac{\partial Z_m}{\partial c}
\end{aligned}
\end{equation} 
where we used the identity $\partial A^{-1} = -A^{-1} A A^{-1}$. The derivative of $Z_m$ varies depending on the role of the capacitor whose value is indicated by $c$. If $c$ is the series capacitor in the matching network ($\in C_s$), then $\frac{\partial Z_m}{\partial c}$ is a matrix with all values equal to zero except the element $\{i,i\}, i \in p$, corresponding to the feeding port equipped with the matching capacitor in series: 
\begin{equation} 
\frac{\partial Z_m}{\partial c}_{ii} = -\frac{1}{\mathrm{i}\omega c^2}.
\end{equation}
If instead $c \in \{C_p,C_t\}$ then
\begin{equation} 
\frac{\partial Z_m}{\partial c} = -Y_m^{-1} \frac{\partial Y_m}{\partial c} Y_m^{-1}.
\end{equation}
If $c$ is a parallel matching capacitor ($\in C_p$), then $\frac{\partial Y_m}{\partial c}$ is a matrix with all values equal to zero, except one element $\{i,i\}, i \in p$, associated with the corresponding feeding port: 
\begin{equation} 
\frac{\partial Y_m}{\partial c}_{ii} = \mathrm{i}\omega.
\end{equation}
If $c$ is a tuning capacitor ($\in C_t$), then the partial derivative is given by
\begin{equation} 
\frac{\partial Y_m}{\partial c} = Y_n^{pl} Y_L^{-1} \left( \frac{\partial Y_L}{\partial c} \right) Y_L^{-1} Y_n^{lp}
\end{equation}
where the partial derivative $\frac{\partial Y_L}{\partial c}$ is a matrix with all the elements equal to zero, except the element $\{i,i\}, i \in l$ corresponding to the lumped elements port equipped with the tuning capacitor:
\begin{equation} 
\frac{\partial Y_L}{\partial c}_{ii} = \mathrm{i}\omega.
\end{equation}

\section*{Conflict of Interest}
The authors have a pending patent application related to the topic of this manuscript (Dkt. No: 046434-0601). AGP is an employee of Q Bio Inc. RL and DKS serve as scientific advisors for the same company.

\section*{Acknowledgements}
This work was supported in part by the Skoltech-MIT Next Generation Program, by NIH R01 EB024536, and by NSF 1453675. It was performed under the rubric of the Center for Advanced Imaging Innovation and Research (CAI$^{\text{2}}$R, www.cai2r.net), a NIBIB Biomedical Technology Resource Center (NIH P41 EB017183).

\bibliographystyle{IEEEtran}
\bibliography{IEEEabrv,References}

\end{document}